\documentclass[aps,preprint,epsfig,rotate]{revtex4}
\begin{document}
\title{Three-particle integrals with Bessel functions.}

 \author{Alexei M. Frolov}
 \email[E--mail address: ]{afrolov@uwo.ca}

\affiliation{Department of pplied Mathematics\\
 University of Western Ontario, London, Ontario N6H 5B7, Canada}

\author{David M. Wardlaw}

\affiliation{Department of Chemistry\\
 Memorial University of Newfoundland, St. John's, Newfoundland and
  Labrador A1C 5S7, Canada}

\date{\today}

\begin{abstract}

Analytical formulas for some useful three-particles integrals are derived. 
Many of these integrals include Bessel and/or trigonometric functions of one 
and two interparticle (relative) coordinates $r_{32}, r_{31}$ and $r_{21}$. 
The formulas obtained in such an analysis allow us to consider three-particle 
integrals of more complicated functions of relative/perimetric coordinates. 
In many actual problems such three-particle integrals can be found in matrix 
elements of the Hamiltonian and other operators. 

\end{abstract}

\maketitle
\newpage

\section{Introduction}

The main goal of this study is to consider some special integrals which are closely
related with various fundamental three- and few-body problems in physics. Possible 
application of such integrals include atomic, molecular and nuclear physics. In many 
of the problems in these areas one finds similar integrals which must be taken over 
three scalar distances and these three distances correspond to the sides (or ribs) 
of the triangle formed by the three `particles'. On the other hand, it is clear now 
that the general theory of three-particle integrals is a rapidly growing area of 
mathematical physics. The methods developed for numerical evaluations of different 
three-particle integrals can be used in various mathematical problems. 

In general, the three-particle (or three-body) integral has the from 
\begin{eqnarray}
 I(\alpha, \beta, \gamma; F) = \int \int \int F(r_{32}, r_{31}, r_{21})
 \exp(-\alpha r_{32} - \beta r_{31} - \gamma r_{21}) r_{32} r_{31} r_{21}
 dr_{32} dr_{31} dr_{21} \label{e1}
\end{eqnarray}
where $\alpha, \beta$ and $\gamma$ are the three real values which are usually
called and considered as the non-linear parameters. The function $F(x,y,z)$ in 
Eq.(\ref{e1}) is an analytical function of each of the three real variables $x, 
y, z$. The generalization of Eq.(\ref{e1}) to the case of complex variables is 
possible, but in this study we do not consider it. Moreover, to simplify the problem 
below we shall assume that the function $F(r_{32}, r_{31}, r_{21})$ does not grow 
`very rapidly' when its arguments increase to the infinity. Such an assumption is 
needed to guarantee the convergence of all arising integrals (see below). The 
variables $r_{ij}$ in Eq.(\ref{e1}) are the three relative coordinates $r_{ij} = 
\mid {\bf r}_i - {\bf r}_j \mid$. These coordinates are scalars and they are 
symmetric upon permutation of their indexes $r_{ij} = r_{ji}$. The relative 
coordinates are not truly independent of each other, since for these coordinates 
the six following conditions must always be obeyed: $r_{ij} \ge \mid r_{ik} - r_{jk} 
\mid$ and $r_{ij} \le r_{ik} + r_{jk}$, where $(i,j,k)$ = (1,2,3).  

The integrals, Eq.(\ref{e1}), can be found in various few-body problems, but 
in Quantum three-body problems they play a central role, since the expressions 
for all matrix elements of the Hamiltonian and overlap matrices are reduced to 
Eq.(\ref{e1}). Furthermore, the explicit expressions for all expectation values 
(regular and singular) also reduce to the computation of formulas each of which
coincides with Eq.(\ref{e1}). Different approaches to analytical and numerical 
computation of the three-body integrals Eq.(\ref{e1}) were developed in the middle
of 1980's \cite{Fro1}, \cite{Fro2}. In particular, in those works it was shown that
the most convenient and simple way to compute such integrals is based on the use
of the three perimetric coordinates $u_1, u_2, u_3$ \cite{Pek}, \cite{MCQ} instead 
of relative coordinates $r_{32}, r_{31}, r_{21}$ mentioned above. The relation 
between these two set of coordinates is linear:
\begin{eqnarray}
 u_1 = \frac12 ( r_{31} + r_{21} - r_{32}) \; \; \; , \; \; \;
 u_2 = \frac12 ( r_{21} + r_{32} - r_{31}) \; \; \; , \; \; \;
 u_3 = \frac12 ( r_{32} + r_{31} - r_{21}) \label{per}
\end{eqnarray}
The inverse relation takes the form $r_{ij} = u_i + u_j$. The use of the three 
perimetric coordinates have a number of advantages in calculations of various 
three-particle integrals. For instance, the three perimetric coordinates $u_1, u_2, 
u_3$ are independent of each other and each of them varies between 0 and $+\infty$. 
The substitution $(r_{32},r_{31},r_{21}) \rightarrow (u_1, u_2, u_3)$ drastically 
simplifies analytical and numerical computations of all three particle integrals. 
In perimetric coordinates $u_1, u_2, u_3$ the basic integral, Eq.(\ref{e1}), is 
written in the form 
\begin{eqnarray}
 I(\alpha, \beta, \gamma; F) = 2 \int_0^{+\infty} \int_0^{+\infty} \int_0^{+\infty} 
 F(u_2 + u_3, u_1 + u_3, u_1 + u_2) \times \nonumber \\
 \exp[-(\alpha + \beta) u_3 - (\alpha + \gamma) u_2 - (\beta + \gamma) u_1] 
 (u_1 + u_2) (u_1 + u_3) (u_2 + u_3) du_1 du_2 du_3 \label{e101}
\end{eqnarray}
where the factor 2 in the front of the integral is the Jacobian of the 
$(r_{32},r_{31},r_{21}) \rightarrow (u_1, u_2, u_3)$ transformation.  
  
The integral, Eq.(\ref{e101}), thus takes the form of a Laplace transformation. It is also 
clear that the image function $I$ will be written in the form $I(\alpha + \beta, \alpha 
+ \gamma, \beta + \gamma; F)$. In general, the calculation of the three-body integrals,
Eq.(\ref{e101}), appears to be closely connected with the Laplace transform. Indeed, 
the tables of Laplace transormations are of great help in analytical and numerical 
computations of various three-body integrals. In our earlier studies we have derived
the explicit formulas for a large number of three-body integrals. Such formulas 
include different regular and singular integrals, integrals with logarithmic terms, etc. 
However, the formulas for some important three particle integrals have never been derived 
in earlier studies. For instance, the integrals which contain one or two Bessel functions 
\cite{FrWa09} and integrals in which one function of relative coordinate is represented in 
some `difficult' forms, e.g., as the infinite power series, or as approximate expansions 
written in terms of other functions. In this work we consider some of such integrals.

\section{Exponential variational expansion}

As mentioned above the three-particle integrals, Eq.(\ref{e1}), arise in various three-body 
problems. In general, all problems related to the construction of highly accurate 
approximations of the actual wave functions for bound states lead to such integrals. This
means that the three-particle integrals, Eq.(\ref{e1}), can be found in all bound state 
problems, including transitions between bound states, time-evolution of bound states, etc. 
On the other hand, analytical/numerical computation of the integrals, Eq.(\ref{e1}), is the 
central part of problems related with the photodetachment and decays of bound states in 
three-body systems. Fortunately, the three-particle integrals, Eq.(\ref{e1}), can be 
considered as the matrix elements between two exponential basis functions written in 
relative coordinates. In general, the exponential variational expansion of the three-body
wave function for the bound $S(L = 0)-$states is written in the form
\begin{eqnarray}
 \Psi = \frac{1}{2} (1 + \kappa \hat{P}_{21}) \sum_{i=1}^{N} C_{i} \exp(-\alpha_{i} r_{32} 
 - \beta_{i} r_{31} - \gamma_{i} r_{21}) \label{eqq1} 
\end{eqnarray}
where $\kappa = \pm 1$ for symmetric systems and $\kappa = 0$ otherwise (see below). The 
generalization of Eq.(\ref{eqq1}) to the case of bound states with arbitrary angular 
momentum $L$ is written in the form \cite{FrEf84}
\begin{eqnarray}
 \Psi = \frac{1}{2} (1 + \kappa \hat{P}_{21}) \sum_{i=1}^{N}
 \sum_{\ell_{1}} C_{i} {\cal Y}_{LM}^{\ell_{1},\ell_{2}} ({\bf r}_{31},
 {\bf r}_{32}) \exp(-\alpha_{i} r_{32} -\beta_{i} r_{31} -\gamma_{i} r_{21}) \label{eqq10} 
\end{eqnarray}
A slightly more complicated (but much more flexible!) generalization of Eq.(\ref{eqq1}) to 
the case of arbitrary $L$ takes the form \cite{Fro01}
\begin{eqnarray}
 \Psi_{LM} = \frac{1}{2} (1 + \kappa \hat{P}_{21}) \sum_{i=1}^{N}
 \sum_{\ell_{1}} C_{i} {\cal Y}_{LM}^{\ell_{1},\ell_{2}} ({\bf r}_{31},
 {\bf r}_{32}) \phi_i(r_{32},r_{31},r_{21}) \exp(-\alpha_{i} u_1 -
 \beta_{i} u_2 - \gamma_{i} u_3) \times \label{eqq2} \\
 \exp(\imath \delta_{i} u_1 + \imath e_{i} u_2
      + \imath f_{i} u_3) \; \; \; , \nonumber
\end{eqnarray}
where $C_{i}$ are the linear (or variational) parameters, $\alpha_i, \beta_i, \gamma_i, 
\delta_i, e_i$ and $f_{i}$ are the real non-linear parameters and $\imath$ is the imaginary 
unit. In the last equations all exponents contain the three perimetric coordinates $u_1, u_2, 
u_3$ instead of relative coordinates $r_{32}, r_{31}, r_{21}$ used in Eq.(\ref{eqq1}). The 
functions ${\cal Y}_{LM}^{\ell_{1},\ell_{2}} ({\bf r}_{31}, {\bf r}_{32})$ in Eqs.(\ref{eqq10}) 
and (\ref{eqq2}) are the bipolar harmonics \cite{Varsh} of the two vectors ${\bf r}_{31} = 
r_{31} \cdot {\bf n}_{31}$ and ${\bf r}_{32} = r_{32} \cdot {\bf n}_{32}$. The bipolar 
harmonics are defined as follows \cite{Varsh}
\begin{equation}
 {\cal Y}_{LM}^{\ell_{1},\ell_{2}} ({\bf x}, {\bf y}) = x^{\ell_{1}}
 y^{\ell_{2}} \sum_{\ell_{1}, \ell_{2}} C^{LM}_{\ell_{1} m_{1};\ell_{2}
 m_{2}} Y_{\ell_{1} m_{1}} ({\bf n}_{x}) Y_{\ell_{2} m_{2}} ({\bf n}_{y})
\end{equation}
where $C^{LM}_{\ell_{1} m_{1};\ell_{2} m_{2}}$ are the Clebsch-Gordan coefficients (see, 
e.g., \cite{Varsh}) and the vectors ${\bf n}_{x} = \frac{{\bf x}}{x}$ and ${\bf n}_{y} = 
\frac{{\bf y}}{y}$ are the corresponding unit vectors constructed for arbitrary non-zero 
vectors ${\bf x}$ and ${\bf y}$. Also, in this equation $L$ is the total angular momentum of 
the three-body system, i.e. $\hat{L}^2 \Psi_{LM} = L (L + 1) \Psi_{LM}$, while $M$ is the 
eigenvalue of the $\hat{L}_z$ operator, i.e. $\hat{L}_z \Psi_{LM} = M \Psi_{LM}$. In actual 
calculations it is possible to use only those bipolar harmonics for which $\ell_{1} + 
\ell_{2} = L + \epsilon$, where $\epsilon = 0$ or 1. The first choice of $\epsilon$ (i.e. 
$\epsilon = 0$) corresponds to the natural spatial parity $\chi_P = (-1)^L$ of the wave
functions \cite{LLQ}. The second choice (i.e. $\epsilon = 1$) represents states with the 
unnatural spatial parity $\chi_P = (-1)^{L+1}$. In almost all works on highly accurate 
bound state computations only the bound states of natural parity are considered. In real 
physical systems only such states are stable. 

The polynomial-type functions $\phi_i(r_{32}, r_{31}, r_{21})$ are used in Eq.(\ref{e1}) 
to represent the inter-particle correlations at short distances. In general, such simple 
polynomial functions allow one to increase the overall flexibility of the variational 
expansion Eq.(\ref{e1}). In many studies, however, these additional functions are chosen in 
the form $\phi_i(r_{32}, r_{31}, r_{21}) = 1$ for $i = 1, \ldots, N$, since the overall
convergence rate of the variational expansion, Eq.(\ref{eqq2}), is already very high.
The operator $\hat{P}_{21}$ in Eq.(\ref{e1}) is the permutation of the identical
particles in symmetric three-body systems, where $\kappa = \pm 1$, otherwise $\kappa = 0$. 

Note that the basis functions, Eq.(\ref{eqq1}), are $(2 L + 1)$-dimensional vectors,
while all matrix elements of the Coulomb three-body Hamiltonian matrix and overlap matrix 
are scalars. In reality, one finds no contradiction here, since the angular integral of the
products of the two and three bipolar harmonics ${\cal Y}_{LM}^{\ell_{1},\ell_{2}} ({\bf 
r}_{31}, {\bf r}_{32})$ equals the products of some linear functions of the $r^2_{31}, 
r^2_{32}, {\bf r}_{31} \cdot {\bf r}_{32}$ variables and $6j-$ and $9j-$symbols, 
respectively (see, e.g., \cite{Efr86} and \cite{FSLA}). The scalar variables $r^2_{31} (=
u^{2}_{3} + u^{2}_{1} + 2 u_1 u_3), r^2_{32} (= u^{2}_{3} + u^{2}_{2} + 2 u_2 u_3), {\bf 
r}_{31} \cdot {\bf r}_{32} = \frac12 (r^2_{32} + r^2_{31} - r^2_{21})$ are easily expressed 
as quadratic functions of the perimetric coordinates. Indeed, by using the relations
between the relative and perimetric coordinates one finds
\begin{eqnarray}
  r^2_{31} = u^{2}_{3} + u^{2}_{1} + 2 u_1 u_3 \; \; \; , \; \; \; 
  r^2_{32} = u^{2}_{3} + u^{2}_{2} + 2 u_2 u_3 \nonumber \\ 
  {\bf r}_{31} \cdot {\bf r}_{32} = \frac12 (r^2_{32} + r^2_{31} - r^2_{21}) = 
  u^2_3 + u_1 u_3 +  u_2 u_3 - u_1 u_2 
\end{eqnarray}
Numerical computations of the corresponding radial integrals are slightly more complicated than 
in the case of $L = 0$. However, all integrals needed in actual bound state computations based 
on the exponential variational expansion, Eq.(\ref{eqq2}), are written in the form of 
Eq.(\ref{e1}), or can be reduced to such a from. This fact explains our permanent interest in 
developing of the new analytical/numerical approaches to calculations of the three-body 
integrals, Eq.(\ref{e1}). 

\section{Integrals of polynomial functions}

First, let us present here our formula for the three-body integral, Eq.(\ref{e1}), which
includes the polynomial function $F(r_{32}, r_{31}, r_{21}) = r^{k}_{32} r^{l}_{31} 
r^{n}_{21}$. In this case the integral is designated as $\Gamma_{k;l;n}(\alpha, 
\beta, \gamma)$ and it is written in the form
\begin{eqnarray}
 \Gamma_{k;l;n}(\alpha, \beta, \gamma) = \int \int \int r^{k}_{32} r^{l}_{31} 
 r^{n}_{21} \exp(-\alpha r_{32} - \beta r_{31} - \gamma r_{21})
 dr_{32} dr_{31} dr_{21} \label{e10}
\end{eqnarray}
where all indexes $k, l, n$ are assumed to be non-negative. In perimetric coordinates
the integral, Eq.(\ref{e10}), is written in the form
\begin{eqnarray}
 \Gamma_{k;l;n}(\alpha, \beta, \gamma) = 2 \int_0^{+\infty} \int_0^{+\infty} 
 \int_0^{+\infty} (u_2 + u_3)^{k} (u_1 + u_3)^{l} (u_1 + u_2)^{n} \times \nonumber \\
 \exp[-(\alpha + \beta) u_3 - (\alpha + \gamma) u_2 - (\beta + \gamma) u_1] 
 du_1 du_2 du_3 \label{e11}
\end{eqnarray}
Analytical evaluation of the integral, Eq.(\ref{e11}), is straightforward. Finally, one finds 
the following formula
\begin{eqnarray}
 && \Gamma_{k;l;n}(\alpha, \beta, \gamma) = 2 \sum^{k}_{k_1=0} \sum^{l}_{l_1=0} 
 \sum^{n}_{n_1=0} C^{k}_{k_1} C^{l}_{l_1} C^{n}_{n_1} 
 \frac{(l-l_1+k_1)!}{(\alpha + \beta)^{l-l_1+k_1+1}}
 \frac{(k-k_1+n_1)!}{(\alpha + \gamma)^{k-k_1+n_1+1}}
 \frac{(n-n_1+l_1)!}{(\beta + \gamma)^{n-n_1+l_1+1}} \nonumber \\
 &=& 2 \cdot k! \cdot l! \cdot n! \sum^{k}_{k_1=0} \sum^{l}_{l_1=0} \sum^{n}_{n_1=0} 
 \frac{C^{k_1}_{n-n_1+k_1} C^{l_1}_{k-k_1+l_1} C^{n_1}_{l-l_1+n_1}}{(\alpha + 
 \beta)^{l-l_1+k_1+1} (\alpha + \gamma)^{k-k_1+n_1+1} (\beta + \gamma)^{n-n_1+l_1+1}} 
 \label{e12}
\end{eqnarray}
where $C^{m}_{M}$ is the number of combinations from $M$ by $m$ ($m$ and $M$ are the
non-negative integers). The formula, Eq.(\ref{e12}), can also be written in a few
other equivalent forms. The function $\frac{n!}{X^{n+1}}$ in Eq.(\ref{e12}) is the 
$A_n(X)$ function introduced by Larson \cite{Lar}. For the first time, One og the authors
produced  the formula, Eq.(\ref{e12}), in the middle of 1980's (see \cite{Fro1}, \cite{Fro2} 
and references therein) for the first time. 

Note that in some of our earlier works the following integral in perimetric coordinates was
considered as the basic three-body integral:
\begin{eqnarray}
 B(a, b, c; p_1, p_2, p_3; q_0, q_1 , q_2, q_3; s) = \int_0^{+\infty} \int_0^{+\infty} 
 \int_0^{+\infty} \frac{u^{p_1}_1 u^{p_2}_2 u^{p_3} \exp(-a u_1 - b u_2 - c u_3)}{(q_0 + 
 q_1 u_1 + q_2 u_2 + q_3 u_3)^{s}} du_1 \times \nonumber \\ 
 du_2 du_3 = \frac{\Gamma(p_1 + 1) 
 \Gamma(p_2 + 1) \Gamma(p_3 + 1)}{\Gamma(s)} \int_0^{\infty} \frac{\exp(-q_0 x) x^{s-1} 
 dx}{(a + q_1 x)^{p_1+1} (b + q_2 x)^{p_2+1} (c + q_3 x)^{p_3+1}} \; \; \; , \; \; \; 
 \label{e21}
\end{eqnarray}
where $\Gamma(x)$ is the Euler's gamma-function (see, e.g., \cite{GR}, \cite{AS}).
This integral depends upon eleven parameters $a, b, c, p_1, p_2, p_3; q_0, q_1 , q_2, q_3$ 
and $s$. Formally, all these parameters must be real and positive (or non-negative). In 
particular, for $s = 1, q_0 = 1$ and $q_1 = q_2 = q_ 3 = 0$ one finds from the last formula 
\begin{eqnarray}
 B(a, b, c; p_1, p_2, p_3; 1, 0, 0, 0; 1) = 
 \frac{\Gamma(p_1 + 1)  \Gamma(p_2 + 1) \Gamma(p_3 + 1)}{a^{p_1+1} b^{p_2+1} c^{p_3+1}}
 \label{e23}
\end{eqnarray}
This formula leads to the generalization of Eq.(\ref{e12}) to the case of non-integer 
values of $P_1, p_2, p_3$. It is often used to operate with the modified basis sets, e.g., with 
the basis set which includes semi-integer powers of perimetric coordinates. 

In some related three-body problems, e.g., scattering, one finds a number of advantages of using 
some non-exponential basis sets, e.g., power-type wave functions of the relative and/or perimetric 
coordinates. In such cases we need to determine different basic integrals. In this study we chose 
not to discuss this interesting problem. Note only the following formula which arises in the case 
of power-type basis functions
\begin{eqnarray}
 G(p_1, p_2, p_3; q_0, q_1 , q_2, q_3; s) = \int_0^{+\infty} \int_0^{+\infty} 
 \int_0^{+\infty} \frac{u^{p_1}_1 u^{p_2}_2 u^{p_3}}{(q_0 + 
 q_1 u_1 + q_2 u_2 + q_3 u_3)^{s}} du_1 du_2 du_3 \nonumber \\ 
 = \frac{\Gamma(p_1 + 1) \Gamma(p_2 + 1) \Gamma(p_3 + 1)}{\Gamma(s) q_1^{p_1+1} 
  q_2^{p_2+1} q_3^{p_3+1}} \cdot \frac{\Gamma(s - p_1 - p_2 - p_3 - 3)}{q_0^{s - p_1 - p_2 
 - p_3 - 3}} \label{e25}
\end{eqnarray}
where it is assumed that $s > p_1 + p_2 + p_3 + 3$ and all values $p_i, q_i$ ($i$ = (1,2,3))
and $q_0$ must be positive.

\section{Derivation of the related integrals}

By using the expression for the $\Gamma_{k;l;n}$ integral, Eq.(\ref{e10}), we can obtain analytical 
formulas for various three-particle integrals. First, consider the matrix elements of the real 
(analytical) functions $f(r_{21})$ and $F(r_{21})$ which are represented by the following series:   
\begin{eqnarray}
 f(r_{32}) = \sum_{n} A_n r^n_{32} \; \; \; \; and \; \; \; \; F(r_{32}) = \sum_{n} 
 A_n r^n_{32} \exp(-B_n r_{32}) \label{e30}
\end{eqnarray}
where the number of terms can be finite, or infinite. The computation of the matrix elements 
of these functions is reduced to the calculation of the two following sums
\begin{eqnarray}
 M_f = \sum_{n} A_n \Gamma_{n+1;1;1}(\alpha, \beta, \gamma) \label{e31}
\end{eqnarray}
and 
\begin{eqnarray}
 M_F = \sum_{n} A_n \Gamma_{n+1;1;1}(\alpha + B_n, \beta, \gamma) \label{e32}
\end{eqnarray}
where the integals $\Gamma_{n;k;l}$ are defined in Eq.(\ref{e1}). If the coefficients $A_n$ in 
these expansions rapidly decrease with $n$, then one needs to compute only a finite number of 
terms in such sums. For instance, if the coefficients in Eq.(\ref{e30}) are the power-type 
functions of some small parameter, then the series in Eq.(\ref{e31}) and Eq.(\ref{e32}) converge 
rapidly. This is the case in various atomic problems related to QED applications, when $A_n 
\sim \alpha^n$, where $\alpha = \frac{e^2}{\hbar c} \approx \frac{1}{137}$ is the dimensionless 
fine-structure constant.

This approach can also be used to produce analytical formulas for more complicated integrals, 
e.g., the general three-body integrals with the Bessel function $j_L(V r_{32})$. First, let us
obtain the computational formula for the following integral 
\begin{eqnarray}
 B^{(0)}_{k;l;n}(\alpha, \beta, \gamma; V) = \int \int \int r^{k}_{32} r^{l}_{31} 
 r^{n}_{21} j_0(V r_{32}) \exp(-\alpha r_{32} - \beta r_{31} - \gamma r_{21})
 dr_{32} dr_{31} dr_{21} \label{e33}
\end{eqnarray}
By using the formula $j_0(x) = \frac{\sin x}{x}$ and Eq.(\ref{e31}) one finds
\begin{eqnarray}
 B^{(0)}_{k;l;n}(\alpha, \beta, \gamma; V) = \sum^{\infty}_{q=0} \frac{(-1)^q V^{2 q}}{(2 q + 1)!} 
 \Gamma_{k + 2 q;l;n}(\alpha, \beta, \gamma) \approx \sum^{q_{max}}_{q=0} \frac{(-1)^q 
 V^{2 q}}{(2 q + 1)!} \Gamma_{k + 2 q;l;n}(\alpha, \beta, \gamma) \label{e34}
\end{eqnarray}
The integral $B^{(0)}_{k;l;n}(\alpha, \beta, \gamma; V)$ in the last equation converges for all 
$V$, but for $V \le 1$ it converges very rapidly. In reality, the maximal value of the index $q$ 
(or $q_{max}$) in Eq.(\ref{e34}) is finite. Numerical investigations indicate that to stabilize 15 
decimal digits for $V \le 1$ one needs to use in Eq.(\ref{e34}) $q_{max}$ = 20 to 40. For $V \ge 
2$ the value of $q_{max}$ rapidly increases up to 50 - 70 and even 100. The same conclusion is 
true about the convergence of the three-body integrals with the Bessel function $j_1(x) = 
\frac{\sin x}{x^2} - \frac{\cos x}{x}$. This integral takes the form
\begin{eqnarray}
 B^{(1)}_{k;l;n}(\alpha, \beta, \gamma; V) = \sum^{\infty}_{q=0} \frac{(-1)^q (2 q + 2) 
 V^{2 q + 1}}{(2 q + 3)!} \Gamma_{k + 2 q + 1;l;n}(\alpha, \beta, \gamma) \label{e35}
\end{eqnarray}
The three-body integrals with the lowest order Bessel functions $j_0(x)$ and $j_1(x)$ are of 
great interest in applications involving the decays and photodetachment of atoms/ions in their 
ground states. The results of numerical computations of some integrals $\Gamma_{k;l;n}(\alpha, \beta, 
\gamma), B^{(0)}_{k;l;n}(\alpha, \beta, \gamma; V)$ and $B^{(1)}_{k;l;n}(\alpha, \beta, \gamma; V)$ 
can be found in Tables I and II. For all integrals with Bessel functions shown in Table II we 
restricted to the accuracy $\approx 1 \cdot 10^{-15}$. To determine the integrals 
$B^{(0)}_{k;l;n}(\alpha, \beta, \gamma; V)$ and $B^{(1)}_{k;l;n}(\alpha, \beta, \gamma; V)$ to such 
an accuracy for $V \le 1$ it was sufficient to use 30 terms in Eqs.(\ref{e34}) and (\ref{e35}). For 
$V = 2$ we used up to 75 terms in these formulas.

The formulas for the three-body integrals $B^{(L)}_{k;l;n}(\alpha, \beta, \gamma; V)$ with other 
spherical Bessel functions $j_L(V r_{32})$ can be obtained by using the same procedure. The result
is
\begin{eqnarray}
 B^{(L)}_{k;l;n}(\alpha, \beta, \gamma; V) = V^{L} \sum^{\infty}_{\kappa=0} 
 \frac{(-1)^{\kappa} V^{2 \kappa} }{2^{\kappa} \kappa! (2 L + 2 \kappa + 1)!!}
 \Gamma_{k + L + 2 \kappa;l;n}(\alpha, \beta, \gamma) 
 \label{e36}
\end{eqnarray}
where $\kappa$ is integer number. In Eq.(\ref{e36}) we have used the following formula for the 
$j_L(x)$ Bessel function
\begin{eqnarray}
 j_L(z) = z^L \sum^{\infty}_{k=0} \frac{(-1)^{k} z^{2 k}}{2^{k} k! (2 L + 2 k + 1)!!} 
 \label{e37}
\end{eqnarray}
Another approach for derivation of these formulas is based on the use of the well-known recursion 
formulas for the spherical Bessel functions \cite{GR} and analytical formulas for the three-body 
integrals containing the $j_0(V r_{32})$ and $j_1(V r_{32})$ Bessel functions, Eqs.(\ref{e34}) and 
(\ref{e35}). The formulas for the integrals containing the spherical Bessel functions of different 
arguments, e.g., $j_L(V r_{31})$ and/or $j_L(V r_{21})$, are easily obtained from the expressions 
for the integrals with the $j_L(V r_{32})$ functions by applying a set of different $\alpha 
\rightarrow \beta \rightarrow \gamma$ substitutions. Here we do not want to produce these formulas, 
since it is rather a technical problem which step-by-step repeats the procedure described above 
for deriving the formulas for the integrals with the $j_0(V r_{32})$ and $j_1(V r_{32})$ Bessel 
functions. Instead we consider in the next Section a more interesting and actual problem which is 
closely related with analytical and numerical computation of other three-body integrals with modified 
Bessel functions $K_0(r), Ki_1(r)$ and $Ki_2(r)$ functions. 

\section{Matrix elements of the Uehling potential}

As is well known (see, e.g., \cite{AB}, \cite{Grei}) in the lowest order approximation the effect 
of vacuum polarisation between two interacting electric charges is described by the Uehling 
potential $U(r)$ \cite{Uehl}. In \cite{FroWa1} we have derived the closed analytical formula for 
the Uehling potential. For atomic systems this formula is written in the following three-term form 
(in atomic units $\hbar = 1, m_e = 1, e = 1$)
\begin{eqnarray}
 U(2 b r) = \frac{2 \alpha Q}{3 \pi} \cdot \frac{1}{r} \Bigl[ 
 \int_1^{+\infty} exp(-2 \alpha^{-1} \xi r) \Bigl(1 + \frac{1}{2 \xi^2}
 \Bigr) \frac{\sqrt{\xi^2 - 1}}{\xi^2} d\xi \Bigr] \nonumber\\
 = \frac{2 \alpha Q}{3 \pi r} \Bigl[ \Bigl(1 + \frac{b^2
 r^2}{3}\Bigr) K_0(2 b r) - \frac{b r}{6} Ki_1(2 b r) - \Bigl(\frac{b^2
 r^2}{3} + \frac{5}{6}\Bigr) Ki_2(2 b r) \Bigr] \; \; \; , \label{Ueh}
\end{eqnarray}
where the notation $Q$ stands for the electric charge of the nucleus, $b = \alpha^{-1}$ and 
$\alpha = \frac{e^2}{\hbar c} \approx \frac{1}{137}$ is the dimensionless fine-structure constant. 
Here and below $\hbar = \frac{h}{2 \pi}$ is the reduced Planck constant (also called the Dirac 
constant), $e$ is the electric charge of the positron and $m_e$ is the mass of the electron 
(= mass of the positron). In Eq.(\ref{Ueh}) $K_0(a)$ is the modified Bessel function of zero order 
(see, e.g, \cite{GR}), i.e.
\begin{eqnarray}
 K_0(z) = \int_0^{\infty} exp(-z \cosh t) dt = \sum_{k=0}^{\infty} 
 (\psi(k+1) + \ln 2 - \ln z) \frac{z^{2 k}}{2^{2 k} (k!)^2} \; \; \; , 
 \label{macd}
\nonumber
\end{eqnarray}
where $\psi(k)$ is the Euler $psi$-function defined by Eq.(8.362) from \cite{GR}. The functions 
$Ki_1(z)$ and $Ki_2(z)$ in Eq.(\ref{Ueh}) are the recursive integrals of the $Ki_0(z) \equiv K_0(z)$ 
function, i.e.
\begin{eqnarray}
 Ki_1(z) = \int_z^{\infty} Ki_0(z) dz \; \; \; , \; \; \; and \; \; \;
 Ki_n(z) = \int_z^{\infty} Ki_{n-1}(z) dz  \; \; \; , \label{repin}
\end{eqnarray}
where $n \ge 1$.

The calculation of the matrix elements of the Uehling potential with the use of our three term 
formula, Eq.(\ref{Ueh}), leads to the following three-body integrals:
\begin{eqnarray}
 K^{(p)}_{k;l;n}(\alpha, \beta, \gamma) = \int \int \int r^{k}_{32} r^{l}_{31} 
 r^{n}_{21} Ki_p(r_{32}) \exp(-\alpha r_{32} - \beta r_{31} - \gamma r_{21})
 dr_{32} dr_{31} dr_{21} \label{Ueh2}
\end{eqnarray}
for $p = 0, 1, 2$ and two other similar integrals which contain $Ki_p(r_{31})$ and $Ki_p(r_{21})$. 
This problem can be solved approximately by using the known analytical formulas for the modified 
Bessel function $K_0(z)$ and for the two lowest recursive integrals of this function. However, the 
overall accuracy of the final solution is not very high. There are some advanced methods which can 
be used to solve this problem, but at this moment we also trying to apply a few alternative 
approaches. One of these methods is based on the original integral representation for the Uehling 
potential, Eq.(\ref{Ueh}). Below, we consider this procedure in detail. 

First, consider the matrix elements of the regular Yukawa-type interparticle potential which is 
written in the form $V_Y(r) = V_0 \frac{exp(-\mu r)}{r}$. The matrix element between the two 
exponential basis functions takes the following form 
\begin{eqnarray}
 V_0 \int \int \int r_{31} r_{21} \exp[-(\alpha + \mu)r_{32} - \beta r_{31} - 
 \gamma r_{21}] dr_{32} dr_{31} dr_{21} = V_0 \Gamma_{0;1;1}(\alpha + \mu, \beta, 
 \gamma)
\end{eqnarray}
in the case of the $V_Y(r_{32})$ interaction. Note also that each of the integrals $\Gamma_{0;1;1}$ 
contains only four terms. This allows one to obtain the following formula in the case of the $U_{21}$ 
potential, which is the (21)-component of the total Uehling potential, Eq.(\ref{Ueh}):
\begin{eqnarray}
 \overline{U}_{21}(2 b \xi) = \int \int \int \frac{exp(-2 b \xi
 r_{21})}{r_{21}} exp(-\alpha r_{32} -\beta r_{31} -\gamma r_{21}) r_{32}
 r_{31} r_{21} dr_{32} dr_{31} dr_{21} = \nonumber \\
 \frac{2}{(\alpha + \beta) (\alpha + \gamma + 2 b \xi) (\beta + \gamma + 2
 b \xi)} \Bigl[ \frac{2}{(\alpha + \beta)^2} + \frac{1}{(\beta + \gamma +
 2 b \xi) (\alpha + \beta)} + \label{Lapl} \\
 \frac{1}{(\alpha + \gamma + 2 b \xi) (\alpha + \beta)} + \frac{1}{(\beta
 + \gamma + 2 b \xi) (\alpha + \gamma + 2 b \xi)} \Bigr] \; \; \; ,
 \nonumber
\end{eqnarray}
where $\alpha + \beta > 0, \alpha + \gamma > 0, \beta + \gamma > 0$ and $\xi > 0$. The first factor 2 
in the numerator is the Jacobian of the linear transformation from relative to perimetric coordinates. 
Analogous expressions can be obtained for the $U(r_{32})$ and $U(r_{31})$ Yukawa-type potentials. In 
fact, such formulas can be derived from Eq.(\ref{Lapl}) simply by performing cyclic permutations of 
three parameters $\alpha, \beta$ and $\gamma$.

The final formula for the matrix elements of the Uehling potential in an arbitrary Coulomb three-body 
system is written in the form 
\begin{eqnarray}
 \frac{2 \alpha}{3 \pi} \int_1^{\infty} \Bigl[ q_1 q_2
 \overline{U}_{21}(2 b \xi) + q_1 q_3 \overline{U}_{31}(2 b \xi) + q_2 q_3
 \overline{U}_{32}(2 b \xi) \Bigr] \Bigl(1 +
 \frac{1}{2 \xi^2} \Bigr) \frac{\sqrt{\xi^2 - 1}}{\xi^2} d\xi \; \; \; ,
 \label{Int51}
\end{eqnarray}
where $q_i$ ($i$ = 1, 2, 3) are the particle charges expressed in atomic units. The expressions for the 
$\overline{U}_{ij}(2 b \xi)$ terms are obtained from  Eq.(\ref{Lapl}). By using the formulas, 
Eqs.(\ref{Lapl}) and (\ref{Int51}), we have developed a number of effective numerical methods for 
accurate evaluation of the three-particle integrals arising in the expansion of the Uehling potential.

\section{Three-particle integrals with two Bessel functions}

A number of actual problems in modern atomic physics lead to three-particle integrals with two 
Bessel functions. For instance, the probability of formation of the negatively charged tritium ion
(T$^-$) during the $(n, {}^3$He; ${}^1$H, ${}^3$H) nuclear reaction in the two-electron ${}^3$He-atom
is reduced to the calculation of the following integral (`probability amplitude' for more details, 
see, e.g., \cite{Fro2013})
\begin{eqnarray}
 & & A_{if} = \int \int \int \Phi_{{\rm T}^{-}}(r_{32}, r_{31}, r_{21}) j_0(V_t \cdot r_{32}) 
 j_0(V_t \cdot r_{31}) \Psi_{{\rm He}}(r_{32}, r_{31}, r_{21}) r_{32} r_{31} r_{21} dr_{32} dr_{31} 
 dr_{21} \nonumber \\ 
 & & = \frac{1}{V^2_t} \int \int \int \Phi_{{\rm T}^{-}}(r_{32}, r_{31}, r_{21})
 sin(V_t \cdot r_{32}) sin(V_t \cdot r_{31}) 
 \Psi_{{\rm He}}(r_{32}, r_{31}, r_{21}) r_{21} dr_{32} dr_{31} dr_{21} \label{eq45}
\end{eqnarray}
where $V_t$ is the speed of the tritium nucleus after the nuclear reaction in the ${}^3$He atom. To
obtain the formula, Eq.(\ref{eq45}), we have used the known fact from atomic physics that the negatively
charged hydrogen ion has only one bound $1^1S(L = 0)-$state. Also, to derive Eq.(\ref{eq45}) we restrict 
ourselves to the case when the incident ${}^3$He atom is in its ground $1^1S(L = 0)-$state. 
 
As mentioned in the second Section the wave functions of the ground $1^1S(L = 0)-$states in the 
two-electron H$^{-}$ ion and He atom are usually approximated with the use of highly accurate variational 
expansion written in the relative/perimetric coordinates. The most advanced of such expansions is the 
exponential variational expansion in relative/perimetric coordinates which takes the following form (for 
the bound $S(L = 0)-$states in the three-body systems):
\begin{eqnarray}
 \psi(r_{32}, r_{31}, r_{21}) = \sum^{N}_{i=1} C_i exp(-\alpha_i r_{32} - \beta_i r_{31} 
 - \gamma_i r_{21}) \label{eq7}
\end{eqnarray} 
where $C_i$ are the variational coefficients and $N$ is the total number of terms in the trial 
function $\psi(r_{32}, r_{31}, r_{21})$. The probability amplitude $A_{if}$ is written as the double 
sum of the following three-particle integrals 
\begin{eqnarray}
 B^{(00)}_{k;l;n}(\alpha,\beta,\gamma) = \int \int \int exp(-\alpha r_{32} - \beta r_{31} 
 - \gamma r_{21}) sin(V_t \cdot r_{32}) sin(V_t \cdot r_{31}) r_{12} dr_{32} dr_{31} dr_{21} 
 \label{eq8}
\end{eqnarray}
The theory of these integrals has not been developed in earlier studies. It was shown in \cite{Fro2013}   
that the integral, Eq.(\ref{eq8}), is reduced to the following double sum (here we apply the 
Cauchy formula) 
\begin{eqnarray}
 B^{(00)}_{0;0;1}(\alpha,\beta,\gamma;V) = \sum^{\infty}_{\kappa=0} \frac{(-1)^{\kappa} V^{2 
 \kappa}}{(2 \kappa + 2)!} \sum^{\kappa}_{\mu=0} C^{2 \mu + 1}_{2 \kappa + 2} 
 \Gamma_{k + 2 \mu + 1;l + 2 \kappa - 2 \mu + 1;n+1}(\alpha, \beta, \gamma) \label{eq46}
\end{eqnarray}
where $C^{k}_{n}$ is the number of combinations from $n$ by $k$ ($n \ge k$) and 
$\Gamma_{k;l;n}(a,b,c)$ is the basic three-particle integral defined above. Note that the integral, 
Eq.(\ref{eq8}), can easily be computed with the use of complex arithmetic. However, such methods
are difficult to use in the general case. In our study to check the formula, Eq.(\ref{eq46}), we have 
used both approaches.  

The integral $B^{(00)}_{0;0;1}(\alpha,\beta,\gamma;V)$ belongs to the new class of three-particle 
integrals $B^{(L_1 L_2)}_{k;l;n}(\alpha,\beta,\gamma;V)$. Such integrals contain the two Bessel
functions $j_L(V r_{32})$ and $j_L(V r_{31})$. In this Section we consider one approach developed
recently for computations of such integrals. First, by using Eq.(\ref{e37}) one
finds the following formula for the product of the two Bessel functions $j_{L_{1}}(a x)$ and 
$j_{L_{2}}(b y)$:
\begin{eqnarray}
 j_{L_{1}}(a x) j_{L_{2}}(b y) = a^{L_{1}} x^{L_{1}} b^{L_{2}} y^{L_{2}} \sum^{\infty}_{p=0}
 \frac{(-1)^p}{2^p p!} \sum^{p}_{q=0} C^{q}_{p} \frac{a^{2q} x^{2 q} b^{2 p - 2 q} y^{2 p - 
 2 q}}{(2 L_1 + 2 q + 1)!! (2 L_2 + 2 p - 2 q + 1)!!} \label{eq47}
\end{eqnarray}
where $p$ and $q$ are both integer non-negative numbers and  $C^{q}_{p}$ is the binomial coefficient. 
In our case we have $a = b = V, x = r_{32}$ and $y = r_{31}$. Therefore, the formula for the 
$B^{(L_1 L_2)}_{k;l;n}(\alpha,\beta,\gamma;V)$ integrals take the form:
\begin{eqnarray}
 & & B^{(L_1 L_2)}_{k;l;n}(\alpha,\beta,\gamma;V) = V^{L_{1} + L_{2}} \sum^{\infty}_{p=0}
 \frac{(-1)^p V^p}{2^p p!} \sum^{p}_{q=0} \frac{C^{q}_{p}}{(2 L_1 + 2 q + 1)!! (2 L_2 + 2 p - 
 2 q + 1)!!} \times \nonumber \\
 & & \Gamma_{k + L_1 + 2 q;l + L_2 + 2 p - 2 q;n}(\alpha, \beta, \gamma)
 \label{eq48}
\end{eqnarray}
This formula is appropriate in all applications where $V \le 10$. For $V \le 1$ the overall 
convergence rate of Eq.(\ref{eq48}) for the $B^{(L_1 L_2)}_{k;l;n}(\alpha,\beta,\gamma;V)$ integrals 
is fast, while for $1 \le V \le 2$ it is relatively fast and for $5 \le V \le 10$ the convergence 
rate can be considered as moderate. In the cases when $V \ge 15 - 20$ one needs to develop some
other methods of computation of the $B^{(L_1 L_2)}_{k;l;n}(\alpha,\beta,\gamma;V)$ integrals, but we
do not pursue this here. 

\section{Three-particle integrals of more complicated functions}

By using the formulas derived above for three-particle integrals with one and two Bessel functions 
we can obtain some useful formulas for more complicated three-particle integrals. In reality, one
finds a large number of three-particle integrals which can be approximated by the integrals with 
Bessel function(s). In this study we restrict ourselves to the consideration of the following 
integral:
\begin{eqnarray}
 J(\alpha, \beta, \gamma; t) = \int \int \int cos\sqrt{r^2_{32} - 2 t r_{32}} \cdot 
 \exp(-\alpha r_{32} - \beta r_{31} - \gamma r_{21}) dr_{32} dr_{31} dr_{21} \label{eq51}
\end{eqnarray}
Such integrals and their $t-$derivatives arise, e.g., in the problem of electron scattering on the 
electric dipole formed by  the two heavy, positively charge particles. The formula for numerical 
evaluation of the $J(\alpha, \beta, \gamma; t)$ integral is written in the form
\begin{eqnarray}
 J(\alpha, \beta, \gamma; t) = \sum^{\infty}_{\kappa=0} \frac{t^{\kappa}}{\kappa!}
 B^{(\kappa-1)}_{k+1;l;n}(\alpha, \beta, \gamma; 1) \label{eq52}
\end{eqnarray}
All terms with $\kappa \ge 1$ in this formula are determined directly with the use of formulas 
given above. Calculations of the term with $\kappa = 0$ contains an additional complication, since
in this case $\kappa - 1 = -1$ and we need to define the integral $B^{(-1)}_{k+1;l;n}(\alpha, \beta, 
\gamma; 1)$. By using the formulas (10.1.11) and (10.1.12) from \cite{AS} one finds $x j_{-1}(x) = 
j_0(x) - x j_{1}(x)$. Therefore, for the $B^{(-1)}_{k+1;l;n}(\alpha, \beta, \gamma; 1)$ integral
we have:  
\begin{eqnarray}
 B^{(-1)}_{k+1;l;n}(\alpha, \beta, \gamma; 1) = B^{(0)}_{k;l;n}(\alpha, \beta, \gamma; 1) - 
 B^{(1)}_{k+1;l;n}(\alpha, \beta, \gamma; 1) \label{eq53}
\end{eqnarray}
Now, we can use the formula, Eq.(\ref{eq52}), to approximate the integral $J(\alpha, \beta, \gamma; t)$
to arbitrarily high, in principle, numerical accuracy. There are many other uses for the integrals 
containing one and two Bessel functions and some functions of relative coordinates $r_{32}, r_{31}$ and 
$r_{21}$. A number of such formulas will be considered in our next study. 

\section{Conclusion}

We have considered the problems of analytical and numerical computation of the three-body (exponential) 
integrals which contain different Bessel functions. For a number of such integrals we have derived 
closed analytical formulas and/or developed effective numerical methods. Our main interest is related 
to the integrals which contain the $j_0(k r_{ij})$ and $j_1(k r_{ij})$ Bessel functions, since such 
integrals play a central role in various problems on photodetachment of the ground $S(L = 0)-$states of 
different atomic and molecular systems. The formulas for the integrals with the spherical Bessel 
functions $j_L(k r_{ij})$ for $L \ge 2$ can be derived by applying the same procedure. The case of 
integrals with the modified Bessel functions, e.g., with the $K_n(b r_{ij})$ functions, is more difficult 
for analytical consideration. However, such integrals are needed in various problems, including derivation
of the closed formulas for the matrix elements of the Uehling potential. 

Derivation of simple analytical formulas for the three-particle integrals with one and two Bessel functions 
has a great value in numerous applications to atomic and nuclear physics. On the other hand, it is a very 
interesting mathematical problem, since the three relative coordinates $r_{32} = r_2, r_{31} = r_1$ and $r_{12}$ 
always form a triangle. In fact, we are dealing with a new class of multiple integrals which have special 
form. Such integrals play a central role in various problems of three- and few-body physics. Note here that
three-particle integrals with one and two Bessel functions are of paramount importance for atomic analysis 
of the species arising during the nuclear $(n;t)-, (n;\alpha)-$ and $(n;p)-$reactions in few electron atoms. 
Another interesting problem is analytical/numerical calculations of the exponential three-body integrals which 
include functions approximated by series explicitly written in terms of the spherical Bessel functions (and 
other Bessel functions). In earlier studies numerical evaluation of such integrlas was a very difficult 
problem which had no effective and reliable solution.  

\begin{center} 
 {\Large Appendix} \\
 {\bf Addition theorem for the spherical Bessel functions} 
\end{center}

Let us discuss here the statement known as the addition theorem for the spherical 
Bessel functions. This theorem plays a great role in the physics of three-body  
systems. On the other hand, it is of interest for the general theory of Bessel 
functions. First, consider the familiar Rayleigh expansion of a plane wave
\begin{eqnarray}
  exp(\imath {\bf k} \cdot {\bf r}_{21}) = \sum^{\infty}_{L=0} \imath^{L} 
  (2 L + 1) j_L(k r) P_L(cos\Theta_{21}) \label{eqa1}
\end{eqnarray}
where $j_L(x)$ is the spherical Bessel function of scalar argument $x$, ${\bf a}
\cdot {\bf b}$ designates the scalar product of the two vectors ${\bf a}$ and 
${\bf b}$, while $P_L(y)$ is the Legendre polynomial. In Eq.(\ref{eqa1}) the 
notation $\Theta_{21}$ stands for the angle between the ${\bf k}$ and ${\bf 
r}_{21}$ vectors.

Now, suppose we have the three-body system (123) and we need to re-write the
Rayleigh expansion, Eq.(\ref{eqa1}), in a different form which contains the plane
waves of the two `new' scalar products ${\bf k} \cdot {\bf r}_{31}$ and ${\bf k} 
\cdot {\bf r}_{32}$ . Such problems always arise in three-body physics. It is 
clear that these two plane waves must uniformly be related to each other. The goal 
of this Appendix is to investigate such a relation in detail. Since the three 
particles in any three-body system always form a triangle, then we can write ${\bf 
r}_{21} = {\bf r}_{31} - {\bf r}_{32}$. Now, one finds from Eq.(\ref{eqa1})
\begin{eqnarray}
 exp(\imath {\bf k} \cdot {\bf r}_{21}) = exp(\imath {\bf k} \cdot {\bf r}_{31})
 exp(-\imath {\bf k} \cdot {\bf r}_{32}) \label{eqa2}
\end{eqnarray} 
By using the Rayleigh expansions twice for the right-hand side of the last 
equation we obtain
\begin{eqnarray}
   \sum^{\infty}_{L=0} \imath^{L} (2 L + 1) j_L(k r_{21}) P_L(cos\Theta_{21}) 
 &=& \sum^{\infty}_{L_1=0} \imath^{L_{1}} (2 L_1 + 1) j_{L_{1}}(k r_{31}) 
 P_{L_{1}}(cos\Theta_{31}) \cdot \sum^{\infty}_{L_2=0} \imath^{L_{2}} \times 
 \nonumber \\
 & & (2 L_2 + 1) j_{L_{2}}(k r_{32}) P_{L_{2}}(-cos\Theta_{32}) \label{eqa3}
\end{eqnarray}
where the notations $\Theta_{31}$ and $\Theta_{32}$ designate the angles between 
the ${\bf k}$ vector and ${\bf r}_{31}$ and ${\bf r}_{32}$ vectors, respectively.

The right-hand side of the last equation can be transformed with the use of the
Cauchy product defined by a discrete convolution as follows:
\begin{eqnarray}
 \sum^{\infty}_{L=0} \Bigl[ \sum^{L}_{\ell=0} \imath^{L} (-1)^{L-\ell} 
 (2 \ell + 1) (2 L - 2 \ell + 1) j_{\ell}(k r_{31}) j_{L-\ell}(k r_{32}) 
 P_{\ell}(cos\Theta_{31}) P_{L-\ell}(cos\Theta_{32}) \Bigr] \label{eqa4}
\end{eqnarray}
It is crucially important here that each of the series in the Rayleigh expansion 
of a plane wave see, Eq.(\ref{eqa1}) and Eq.(\ref{eqa2})) converge absolutely
\cite{Rudin}. From here and Eq.(\ref{eqa3}) we can write
\begin{eqnarray}
 (2 L + 1) j_L(k r_{21}) P_L(cos\Theta_{21}) &=& (-1)^L \sum^{L}_{\ell=0} 
 (-1)^{\ell} (2 \ell + 1) (2 L - 2 \ell + 1) j_{\ell}(k r_{31}) 
 j_{L-\ell}(k r_{32}) \times \nonumber \\
 & & P_{\ell}(cos\Theta_{31}) P_{L-\ell}(cos\Theta_{32}) \label{eqa5}
\end{eqnarray}
By multiplying both sides of the last equation by $P_L(cos\Theta_{21})$
and performing the integration over the $\Theta_{21}, \Theta_{32}$ and $\phi_{32}$ 
angles (or over all possible orientations of the ${\bf r}_{32}$ vector) we find the 
following expression 
\begin{eqnarray}
 j_L(k r_{21}) &=& \frac{(-1)^{L}}{4} \sum^{L}_{\ell=0} 
 (-1)^{\ell} (2 \ell + 1) (2 L - 2 \ell + 1) j_{\ell}(k r_{31}) 
 j_{L-\ell}(k r_{32}) \times \nonumber \\
 & & \int_{-\frac{\pi}{2}}^{\frac{\pi}{2}} \int_{-\frac{\pi}{2}}^{\frac{\pi}{2}} 
 P_{\ell}(cos\Theta_{31}) P_{L-\ell}(cos\Theta_{32}) 
 P_L(cos\Theta_{21}) \sin\Theta_{21} d\Theta_{21} \sin\Theta_{32} d\Theta_{32} 
 \; \; \; . \label{eqa51}
\end{eqnarray}
This formula contains the spherical Bessel functions $j_L(k r_{21})$ written 
in terms of the spherical Bessel functions $j_n(k r_{32})$ and $j_n(k r_{31})$ 
of two other arguments, where the three vectors ${\bf r}_{32}, {\bf r}_{31}, 
{\bf r}_{21}$ form a triangle, i.e. ${\bf r}_{32} + {\bf r}_{21} = {\bf r}_{31}$. 
As it follows from the definition of the $cos\Theta_{ij}$ angles we always have 
$\Theta_{21} + \Theta_{31} + \Theta_{32} = \pi$. The formula, Eq.(\ref{eqa51}), 
is known as the addition theorem for the spherical Bessel functions.

\newpage
  \begin{table}[tbp]
   \caption{Numerical values of some three-body integrals $\Gamma_{k;l;n}(\alpha, \beta, \gamma)$.}
     \begin{center}
     \begin{tabular}{| c | c | c | c | c | c | c | c | c |}
      \hline\hline
 $k$ & $l$ & $m$ & $\alpha$ & $\beta$ & $\gamma$ &  $\Gamma_{k;l;n}(\alpha, \beta, \gamma)$ & $\gamma$ & 
 $\Gamma_{k;l;n}(\alpha, \beta, \gamma)$ \\
     \hline
 0 & 2 & 1 & 2.35 & 1.41  & 0.567 & 0.132484880489827E+00 & -0.567 & 0.484535355001714E+01 \\

 1 & 2 & 1 & 2.35 & 1.41  & 0.567 & 0.105479781157007E+00 & -0.567 & 0.462617958966529E+01 \\

 2 & 2 & 1 & 2.35 & 1.41  & 0.567 & 0.123759737118974E+00 & -0.567 & 0.683620356276100E+01 \\

 3 & 2 & 1 & 2.35 & 1.41  & 0.567 & 0.190628938378487E+00 & -0.567 & 0.138242778966704E+02 \\
 
 4 & 2 & 1 & 2.35 & 1.41  & 0.567 & 0.362095286177389E+00 & -0.567 & 0.356816617385975E+02 \\

 5 & 2 & 1 & 2.35 & 1.41  & 0.567 & 0.815657409095427E+00 & -0.567 & 0.112342033402992E+03 \\

 6 & 2 & 1 & 2.35 & 1.41  & 0.567 & 0.212162348108085E+01 & -0.567 & 0.417926993577783E+03 \\

 7 & 2 & 1 & 2.35 & 1.41  & 0.567 & 0.625059393550668E+01 & -0.567 & 0.179435469496013E+04 \\

 8 & 2 & 1 & 2.35 & 1.41  & 0.567 & 0.205551903374530E+02 & -0.567 & 0.873301210942717E+04 \\

 9 & 2 & 1 & 2.35 & 1.41  & 0.567 & 0.745934650018583E+02 & -0.567 & 0.475056243580342E+05 \\
  \hline\hline
  \end{tabular}
  \end{center}
  \end{table}
  \begin{table}[tbp]
   \caption{Numerical values of some three-body integrals $B^{(0)}_{k;l;n}(\alpha, \beta, \gamma; V)$ 
            and $B^{(1)}_{k;l;n}(\alpha, \beta, \gamma; V)$.}
     \begin{center}
     \begin{tabular}{| c | c | c | c | c | c | c | c | c |}
      \hline\hline
 $k$ & $l$ & $m$ & $\alpha$ & $\beta$ & $\gamma$ & V & $B^{(0)}_{k;l;n}(\alpha, \beta, \gamma; V)$ & 
 $B^{(1)}_{k;l;n}(\alpha, \beta, \gamma; V)$ \\
     \hline
 3 & 2 & 1 & 2.35 & 1.41 & 0.567 & 0.25 & 0.18233241643012516E+00 & 0.290930992106451E-01 \\ 

 5 & 2 & 1 & 2.35 & 1.41 & 0.567 & 0.25 & 0.75291471135429875E+00 & 0.166432412830887E+00 \\ 

 3 & 2 & 1 & 2.35 & 1.41 & 0.567 & 0.50 & 0.15968050735256670E+00 & 0.522255954684081E-01 \\

 5 & 2 & 1 & 2.35 & 1.41 & 0.567 & 0.50 & 0.59041249572520414E+00 & 0.278021233893212E+00 \\

 3 & 2 & 1 & 2.35 & 1.41 & 0.567 & 1.00 & 0.94868174980045456E-01 & 0.691516883096556E-01 \\ 

 5 & 2 & 1 & 2.35 & 1.41 & 0.567 & 1.00 & 0.20605506256710767E+00 & 0.274928833359198E+00 \\ 

 3 & 2 & 1 & 2.35 & 1.41 & 0.567 & 1.50 & 0.40374337963233781E-01 & 0.554457473644749E-01 \\

 5 & 2 & 1 & 2.35 & 1.41 & 0.567 & 1.50 & 0.20605506256710767E+00 & 0.274928833359198E+00 \\

 3 & 2 & 1 & 2.35 & 1.41 & 0.567 & 2.00 & 0.11173049407361310E-01 & 0.340384106321226E-01 \\

 5 & 2 & 1 & 2.35 & 1.41 & 0.567 & 2.00 & -0.35522376544132919E-01 & 0.316198754574614E-01 \\
  \hline\hline
  \end{tabular}
  \end{center}
  \end{table}

\begin{thebibliography}{10}

\bibitem{Fro1}A.M. Frolov, {\it Highly accurate variational solutions for the
Coulomb three-body problem}. Preprint IAE-4274/12, 21 p. (1986) (in Russian)
(1986), unpublished.

\bibitem{Fro2}A.M. Frolov, Sov. Phys. JETP {\bf 92}, 1100 (1987) [Zh. Eksp. 
Teor. Fiz. {\bf 92}, 1959 (1987)].

\bibitem{Pek}C.L. Pekeris, Phys. Rev. {\bf 112}, 1649 (1958).

\bibitem{MCQ}R. McWeeny and B.T. Sutcliffe, \textit{Methods of Molecular 
Quantum Mechanics} (Academic Press, New York, (1969)), Chp. 7.

\bibitem{FrWa09} A.M. Frolov and D.M. Wardlaw, Phys. Rev. A, {\bf 79}, 032703 
(2009). 

\bibitem{FrEf84}A.M. Frolov and V.D. Efros, JETP Letters, {\bf 39}, 544 (1984) 
[Pis'ma Zh. Eksp. Teor. Fiz. {\bf 39}, 449 (1984)].

\bibitem{Fro01}A.M. Frolov, Phys. Rev. E {\bf 64}, 036704 (2001).

\bibitem{Varsh} D.A. Varshalovich, A.N. Moskalev and V.K. Khersonskii, {\it 
Angular Momentum in Quantum Mechanics. Non-Relativistic Theory}, 3rd. edn. 
(Oxford, England, Pergamonn Press (1977)).

\bibitem{LLQ}L.D. Landau and E.M. Lifshitz, {\it Quantum Mechanics. 
Non-Relativistic Theory}, 3rd. edn. (Oxford, England, Pergamonn Press (1977)).

\bibitem{Efr86}V.D. Efros, Sov. Phys. JETP, {\bf 63}, 5 (1986) [Zh. Eksp. Teor. 
Fiz. {\bf 90}, 10 (1986)].

\bibitem{FSLA} A.M. Frolov and V.H. Smith, Jr., Phys. Rev. A, {\bf 53}, 3853 
(1996).

\bibitem{Lar}S. Larsson, Phys. Rev. {\bf 169}, 49 (1968).

\bibitem{GR}I.S. Gradstein and I.M. Ryzhik, \textit{Tables of Integrals,
Series and Products}, (6th revised ed., Academic Press, New York, (2000)).

\bibitem{AS}\textit{Handbook of Mathematical Functions}, edited by M.
Abramowitz and I.A. Stegun, (Dover, New York, (1972)).

\bibitem{AB}A.I. Akhiezer and V.B. Beresteskii, {\it Quantum
Electrodynamics}, (4th Ed., Nauka (Science), Moscow (1981)), Chps. 4 and 5
(in Russian).

\bibitem{Grei}W. Greiner and J. Reinhardt, {\it Quantum Electrodynamics.}
(4th. Ed., Springer Verlag, Berlin, (2010)).

\bibitem{Uehl}E.A. Uehling, Phys. Rev. {\bf 48}, 55 (1935).

\bibitem{FroWa1}A.M. Frolov and D.M. Wardlaw, Eur. Phys. Jour. B {\bf 63}, 
339 (2012).

\bibitem{Fro2013}A.M. Frolov, Eur. Phys. J. D {\bf 67}, 126 (2013).

\bibitem{Rudin} W. Rudin, \textit{Principles of the Mathematical Analysis}
(McGraw-Hill Book Company, New York, (1964)), Chp. 3.

\end{thebibliography}
\end{document}